# Red Luminescence in H-doped β-Ga$_2$O$_3$

Thanh Tung Huynh[1], Ekaterine Chikoidze[2], Curtis P. Irvine[1], Muhammad Zakria[1], Yves Dumont[2], Ferechteh H. Teherani[3], Eric V. Sandana[3], Philippe Bove[3], David J. Rogers[3], Matthew R. Phillips[1], Cuong Ton-That[1]*

[1] *School of Mathematical and Physical Sciences, University of Technology Sydney, Ultimo, NSW 2007, Australia*

[2] *Groupe d'Etude de la Matière Condensée (GEMaC), Université de Versailles Saint Quentin – CNRS, Université Paris-Saclay, 45Av. des Etats-Unis, 78035 Versailles Cedex, France*

[3] *Nanovation, 8 route de Chevreuse, 78117 Chateaufort, France*

* Email: cuong.ton-that@uts.edu.au

**Abstract**

The effects of hydrogen incorporation into β-Ga$_2$O$_3$ thin films have been investigated by chemical, electrical and optical characterization techniques. Hydrogen incorporation was achieved by remote plasma doping without any structural alterations of the film; however, X-ray photoemission reveals major changes in the oxygen chemical environment. Depth-resolved cathodoluminescence (CL) reveals that the near-surface region of the H-doped Ga$_2$O$_3$ film exhibits a distinct red luminescence (RL) band at 1.9 eV. The emergence of the H-related RL band is accompanied by an enhancement in the electrical conductivity of the film by an order of magnitude. Temperature-resolved CL points to the formation of abundant H-related donors with a binding energy of 28 ± 4 meV. The RL emission is attributed to shallow donor–deep acceptor pair recombination, where the acceptor is a $V_{Ga}$-H complex and the shallow donor is interstitial H. The binding energy of the $V_{Ga}$-H complex, based on our experimental considerations, is consistent with the computational results by Varley *et al.* [J. Phys.: Condens. Matter, 23, 334212, 2011].

# I. INTRODUCTION

$Ga_2O_3$ has attracted great interest in recent years due to its prospects for use in next generation high-power electronics, deep-ultraviolet optoelectronics, radiation detection and gas sensing devices.[1,2] The most stable β-phase of $Ga_2O_3$ possesses a high electrical breakdown field (~ 8 MV/cm), which is greater than both GaN and SiC currently being used in state-of-the-art high power electronic devices.[1,2] β-$Ga_2O_3$ typically exhibits n-type conductivity; however, the question remains as to whether this unintentional conductivity is due to impurities and/or native defects. Density functional theory calculations predict that both interstitial hydrogen ($H_i$) and hydrogen trapped at oxygen vacancies ($H_O$) act as shallow donors.[3,4] This behavior is unexpected and different from the behaviour of hydrogen in III-nitrides, in which it acts only as a compensating centre and always counteracts the prevailing conductivity.[5] $Ga_2O_3$ is typically doped with hydrogen by ion implantation, annealing in molecular $H_2$ gas or direct plasma exposure at temperatures above 350°C. Pearton's group demonstrated that hydrogen incorporation in β-$Ga_2O_3$ produced an IR absorption peak at 3437 $cm^{-1}$, assigned to the $V_{Ga}$-2H defect complex.[6,7] Investigations using X-ray photoemission showed that hydrogen termination causes downward band bending and an associated surface accumulation of electrons in β-$Ga_2O_3$ crystals,[8] which is in agreement with observed increases in the electrical conductivity of $Ga_2O_3$ films after hydrogen absorption in gas sensing devices.[9] However, this result does not corroborate with the data from current transient spectroscopy, which reveals that hydrogen profoundly reduces the concentration of shallow donors responsible for n-type conduction.[10] Unlike the electronic properties, which have been extensively explored, there is a distinct lack of information on optical signatures of hydrogen in $Ga_2O_3$.

Luminescence in the ultraviolet-visible spectral range from 1.8 to 3.8 eV has been reported for $\beta$-$Ga_2O_3$ bulk and nanostructures; however, the observed emission bands are highly dependent on the sample growth conditions and sample morphology.[5,11-13] Computational and Electron Paramagnetic Resonance (EPR) studies demonstrated that self-trapped holes are thermally stable in $\beta$-$Ga_2O_3$,[14-16] and hence serve as a precursor for the formation of self-trapped excitons (STEs), which have been proposed as responsible for the strong UV emission in bulk $\beta$-$Ga_2O$.[17,18]. Zhou *et al.* [13] reported a red luminescence (RL) band 1.78 eV with a short recombination time (< 50 ns) in nanowire-like structures and attributed this emission to an amorphous $\beta$-$Ga_2O_3$ shell on the nanowires. On the other hand, the RL centred at ~ 2.0 eV in $Ga_2O_3$ nanosheets has been assigned to donor-acceptor-pair (DAP) recombination involving nitrogen acceptors.[19] However, in spite of reports of RL in $\beta$-$Ga_2O_3$, these results are equivocal because of interference from unintentional impurities and the polycrystalline nature of the samples. Here, we report the characteristics of the RL in H-doped $\beta$-$Ga_2O_3$ films. Based on the combination of chemical, electrical and optical studies, the results present unambiguous evidence for the origin of the RL in $\beta$-$Ga_2O_3$.

## II. EXPERIMENTAL DETAILS

Nominally undoped $Ga_2O_3$ thin films (350 nm thick) were grown on *c*-sapphire substrates by pulsed laser deposition (PLD) using a Coherent LPX KrF ($\lambda$ = 248 nm) laser as described elsewhere.[20] Elemental analysis by Glow Discharge Optical Emission Spectroscopy revealed the presence of only Ga and O, and no evidence of the common donor impurities Si and Sn incorporated in the film.[20] Hydrogen was incorporated into the film by remote plasma treatment using an RF plasma generator (100 W power, 0.5 torr hydrogen pressure). The plasma treatment was performed for 40 mins at 200$^o$C; the sample was kept in $H_2$ gas during heating and cooling cycles in order to suppress the inadvertent out-diffusion of plasma-induced impurities. The films were analysed by X-ray Diffraction (XRD) using a Bruker D8 Discover diffractometer and Atomic Force Microscopy (AFM) with a Park XE7 operating in non-contact mode. Electrical measurements were conducted using a van der Pauw configuration in a custom-built high-impedance analysis system (all samples were $0.5 \times 0.5$ cm$^2$ in size.) Due to strong inhomogeneities with depth for the H-doped film, only sheet resistance values are meaningful and reported in this work. X-ray Photoelectron Spectroscopy (XPS) data were collected at a photon energy of 1486 eV on the Soft X-ray Spectroscopy beamline at the Australian Synchrotron. Raman spectroscopy was conducted in backscattering geometry using a Horiba Jobin Yvon LabRAM HR800 spectrometer with 458 nm laser excitation line. The optical properties of the films were characterized by cathodoluminescence (CL) spectroscopy using an FEI Quanta 200 scanning electron microscope (SEM) equipped with a parabolic mirror collector and an Ocean Optics QE65000 spectrometer. For temperature-dependent CL spectroscopy, the sample was mounted on the cryostat cold stage, which enables CL measurements at both low and elevated temperatures. The beam current was varied by changing the condenser lens setting of the SEM and measured by a Faraday cup on the sample holder. All CL spectra were corrected for the total system response of the light collection system. To confirm the reproducibility of film properties, the plasma treatment and CL measurements were repeated twice on different samples.

### III. RESULTS AND DISCUSSION

The XRD 2θ/ω pattern for the film shows three diffraction peaks at 2θ = 18.73$^0$, 38.22$^0$, and 58.87$^0$, which are indexed as the (-201), (-402) and (-603) reflections of monoclinic β-Ga$_2$O$_3$ (Fig. 1). These XRD peak positions and their relative intensities are well matched with bulk crystals and consistent with (-201) oriented β-Ga$_2$O$_3$ bulk crystals.[21] A typical ω rocking curve is shown in the inset, revealing a full width at half maximum (FWHM) of 0.21º, comparable with the reported value for PLD grown (-201) β-Ga$_2$O$_3$ films on a *c*-sapphire substrate.[22] A Tauc plot analysis of optical absorption data for direct allowed transitions, shown in Supplemental Material Fig. S1, yields an optical bandgap of 4.8 eV, also consistent with β-Ga$_2$O$_3$ films grown on a *c*-plane sapphire substrate.[23] Raman spectra for the undoped and H-doped β-Ga$_2$O$_3$ films, shown in Fig. S2, exhibit five peaks as expected for the dominant vibrational modes of β-Ga$_2$O$_3$.[23,24] The peaks are sharp and narrow indicating a high crystalline quality of the film. No Raman peaks belonging to other Ga$_2$O$_3$ polymorphs and no changes to the Raman modes as a result of the plasma treatment were detected. This result is consistent with single phase β-Ga$_2$O$_3$ in both the undoped and H-doped Ga$_2$O$_3$ films.

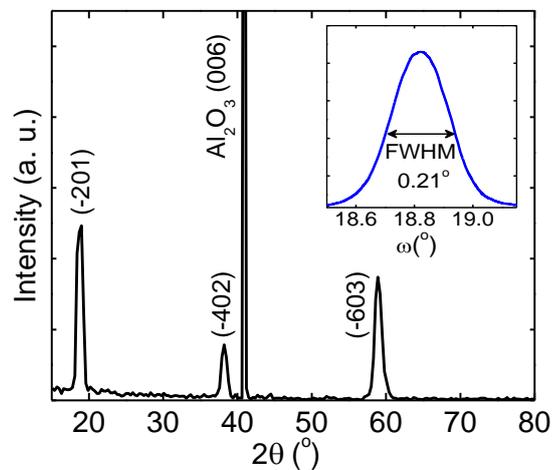

**Fig. 1.** XRD 2θ/ω pattern for the Ga$_2$O$_3$ film grown on a *c*-plane sapphire substrate, showing three diffraction peaks corresponding to the (-201), (-402) and (-603) reflections of (-201) oriented β-Ga$_2$O$_3$. Inset: ω rocking curve for the (-201) XRD peak with a FWHM of 0.21°.

AFM images for the as-grown and H-doped β-Ga$_2$O$_3$ films in Fig. 2(a) show a granular surface structure, similar to those reported by other authors.[25] The as-grown film has an RMS roughness of 6.4 ± 0.4 nm and an average grain size below 100 nm. The H-doped film exhibits a similar morphology with a 6.9 ± 0.4 nm RMS roughness, indicating that the film morphology was little affected by the plasma treatment. Previous SIMS analysis of β-Ga$_2$O$_3$ single crystals doped with deuterium under similar plasma conditions revealed that deuterium is incorporated to a depth of ~200 – 400 nm.[26] Due to the semi-insulating nature of these films, electrical measurements were performed at room and elevated temperatures; however, the samples were kept below 450 K to eliminate the possibility of hydrogen out-diffusion.[26] The incorporation of hydrogen into the β-Ga$_2$O$_3$ film leads to a decrease in sheet resistance by ~ 10 times, as shown in Fig. 2(b). The sheet resistance at 300 K was observed to drop from $10^{10}$ Ω/sq to $4 \times 10^8$ Ω/sq after the H incorporation. This conductivity increase is consistent with surface electron accumulation observed in β-Ga$_2$O$_3$ crystals.[8]

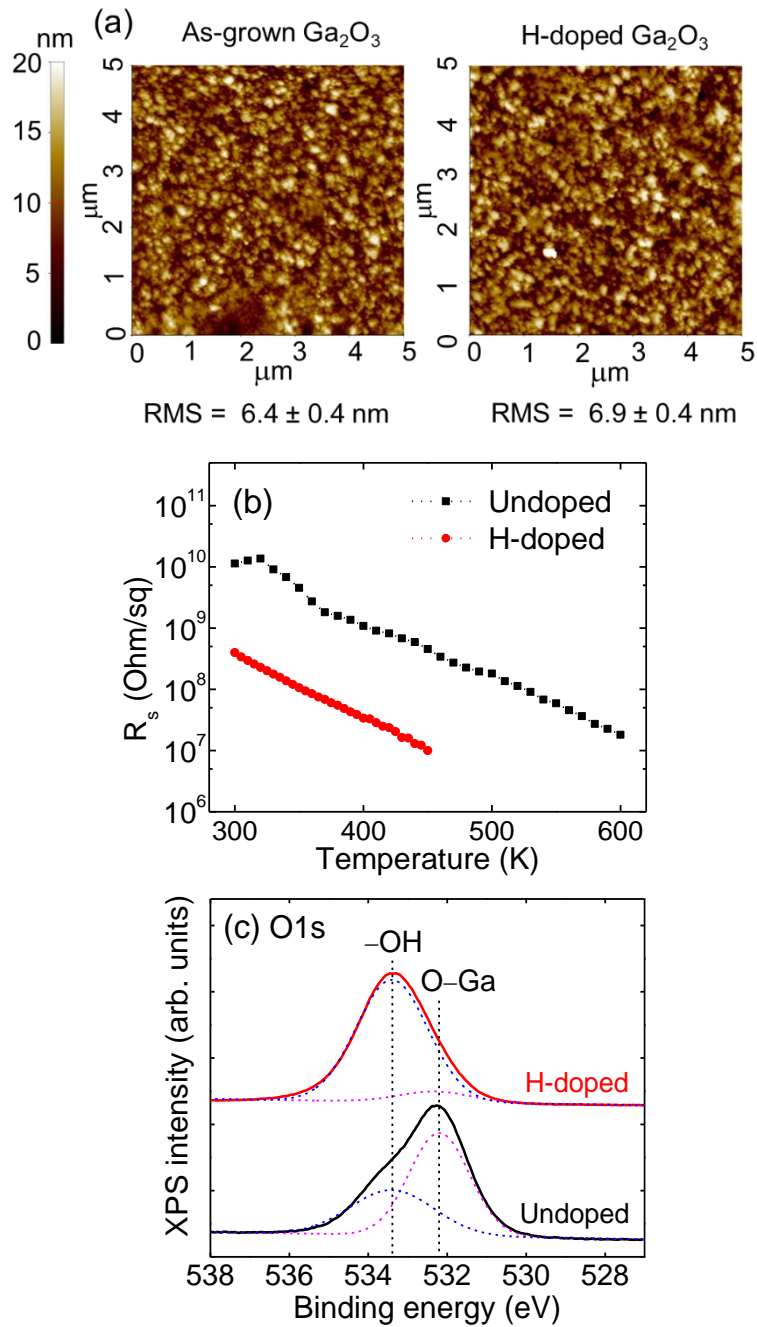

**Fig. 2.** (a) AFM images of the β-Ga$_2$O$_3$ film before and after the H doping. The RMS roughnesses are 6.4 and 6.9 nm for the as-grown and H-doped films, respectively. (b) Sheet

resistance versus temperature for the as-grown (undoped) and H-doped $Ga_2O_3$ films, showing an increase of the electrical conductivity by ~ 10 times due to the H doping. (c) O 1s XPS spectra for the undoped and H-doped $Ga_2O_3$ films showing a binding energy shift from 532.3 to 533.4 eV resulting from the capture of H by O atoms in the near-surface region. The spectra can be fitted with two Voigt functions denoted as O–Ga and –OH; the dashed curves are the theoretical fits.

In order to investigate changes in the surface electronic structure due to H incorporation, the oxygen bonding in the films was analysed via core level XPS analysis using synchrotron X-rays. All survey spectra reveal the presence of Ga, O and adventitious C without any other discernible peaks. The O $1s$ spectra prior to and after H doping are displayed in Fig. 2(c). The O 1s level exhibits a dramatic shift to a higher binding energy as well as changes in spectral shape from an asymmetrical board peak to a narrow peak after H doping. The spectra can be deconvoluted into two Voigt functions with a linear background. The main component at 532.3 eV labelled "O–Ga" in Fig. 2(c) is associated with O atoms bonded to Ga in the $Ga_2O_3$ lattice. The second component, which is located at 1.1 eV higher in binding energy and denoted as "–OH", can be assigned to O–H bonds.[8,27] The remote H plasma doping turns the film surface into a higher valency oxide of Ga with the –OH component completely dominant in the H-doped spectrum. This suggests that plasma-induced H radicals are absorbed into β-$Ga_2O_3$ with high efficiency and form strong bonds with O atoms, which are responsible for the dramatic increase in the film conductivity shown in Fig. 2(b). These results are consistent with the theoretical prediction that lone pairs of the threefold coordinated O atoms in β-$Ga_2O_3$ can efficiently capture $H_i$ without influencing the $Ga_2O_3$ lattice.[4]

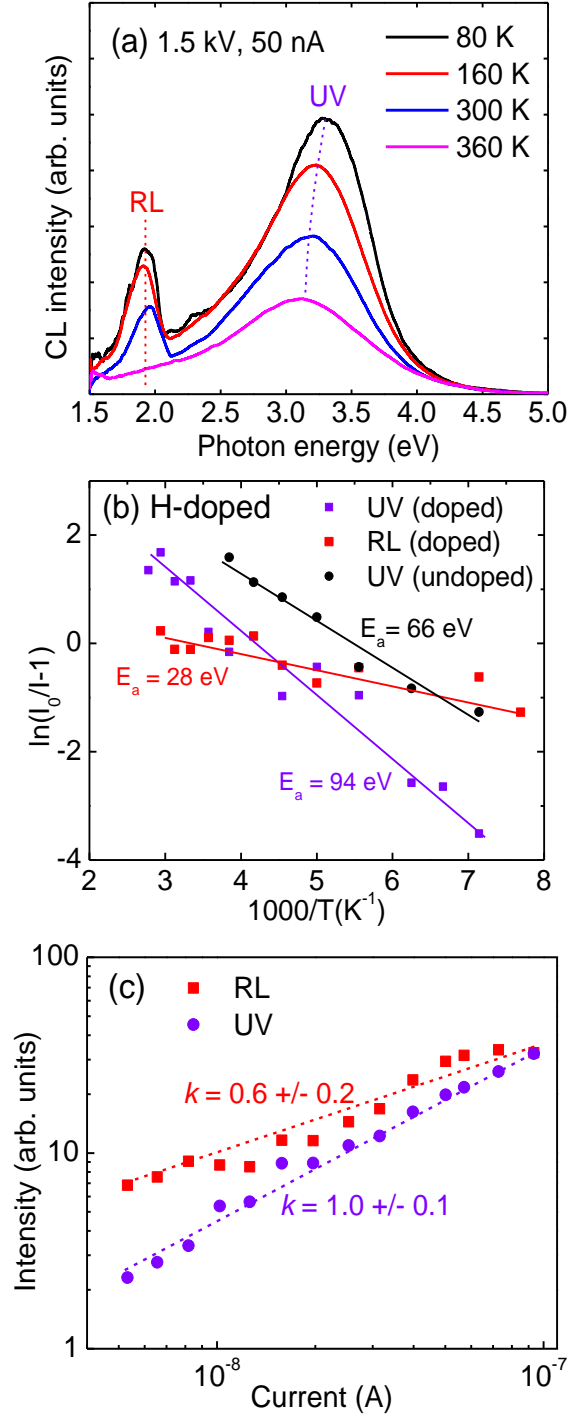

**Fig. 3.** (a) Temperature-resolved CL spectra for H-doped β-Ga$_2$O$_3$ showing two emission bands: a UV one at ~ 3.3 eV and a RL at ~ 1.9 eV. The RL is completely quenched at $T > 340$ K. (b) Arrhenius analysis of the UV and RL integrated intensities yielding activation energies: $E_a$(UV) = 94 ± 7 meV and $E_a$(RL) = 28 ± 4 meV for the H-doped β-Ga$_2$O$_3$, and $E_a$(UV) = 66 ± 4 meV for the undoped β-Ga$_2$O$_3$. (c) Dependence of UV and RL intensities on excitation beam current with $E_b$ = 1.5 keV and $T$ = 80 K. A power-law ($I_{CL} \propto I_B^k$) fit reveals a linear dependence of the UV band and sub-linear dependence of the RL band with $k$ (RL) = 0.6 ± 0.2.

Figure 3(a) shows the temperature-resolved CL spectra for the H-doped β-Ga$_2$O$_3$ film, acquired at 1.5 kV (corresponding to a sampling depth of ~30 nm) which reveal a distinct RL band at 1.9 eV originating from the near-surface region where H dopants are most abundant. The CL emission properties are uniform across the film surface. The overall UV peak of the H-doped film is slightly red shifted from 3.29 eV at 80 K to 3.12 eV at 360 K, probably arising from the overlap with the enhanced defect-related blue emission due to the plasma treatment.[11] Compared with the spectra for the undoped β-Ga$_2$O$_3$ film (Fig. S3), the thermal behaviour of the UV emission band, originating from STEs in Ga$_2$O$_3$,[17] is almost identical, while the RL is completely absent in the undoped β-Ga$_2$O$_3$. This observation is in agreement with the bonding configuration predicted for hydrogen in β-Ga$_2$O$_3$ where the capture of H$_i$ by O lone pairs has little effect on the crystal lattice,[4] which, in turn, would not impact the behaviour of STEs. For comparison, CL spectra were also acquired at acceleration voltage $V_b$ = 4 and 7 kV (Fig. S4), corresponding to a sampling depth of ~150 and 390 nm, respectively, where the H doping concentration is at least one order of magnitude lower than in the near-surface region.[26] It is clear from Fig. S4 that while the energy of the STE emission in the H-doped film remains invariant at 3.24 eV with increasing sampling depth, the RL is present only in the heavily H-doped near-surface layer. The broadening of the 3.4 eV emission at 1.5 kV could arise from plasma-induced potential fluctuations in the near-surface region. The origin of the RL was investigated further by examining the films treated in Ar and N$_2$ plasmas under identical remote plasma conditions (Fig. S5). It is clear that the RL is non-existent in the Ar or N$_2$ plasma-treated films, ruling out the notion that this emission might arise from the near-surface amorphous layer or recombination involving N-related acceptor states in β-Ga$_2$O$_3$.[13,19] The RL was found to quench quickly with increasing temperature and completely disappear at temperatures above 340 K. Arrhenius analysis of the UV and RL bands yields activation energies of: $E_a$ (UV) = 94 ± 7 meV and $E_a$ (RL) = 28 ± 4 meV for the H-doped β-$Ga_2O_3$, and $E_a$ (UV) = 66 ± 4 meV for the as-grown β-$Ga_2O_3$. The increase in the activation energy for the STE emission after the H doping could be due to a slight shift in the position of the O atom that capture $H_i$, as predicted by theoretical calculations.[8] This leads to alteration in the potential well that localises the hole in the STE. The activation energy of 28 meV for the RL is within the binding energy range of 17 – 30 meV that has been measured for Si and Ge shallow donors in β-$Ga_2O_3$.[28,29] Combined with the electrical results shown in Figure 2, this suggests that the activation of the RL and the high conductivity of the H-doped film are both related to the ionization energy of a H-related shallow donor, which is formed by the capture of interstitial H by O atoms.[3] At low temperatures (and below ~320 K as revealed by the temperature-resolved CL), the electron remains bound to the H-related donor and, as described below, participates in DAP recombination with a deep acceptor, giving rise to the RL emission. At higher temperatures, the electron becomes delocalized, resulting in the rapid decrease in the DAP intensity as shown in Figure 3. While hydrogen has been shown to be an effective n-type dopant in β-$Ga_2O_3$ to achieve high-conductivity films for use in hydrogen sensing, the binding energies of hydrogen-related dopants have not yet been experimentally determined. Our results are, however, qualitatively supported by experimental investigations which found that muonium (an exotic atom with similar electronic levels to hydrogen in semiconductors) is a shallow donor in β-$Ga_2O_3$ with a binding energy between 15 and 30 meV.[30]

To investigate the luminescence bands in more detail, excitation power-dependent CL measurements for the H-doped $Ga_2O_3$ films was conducted by varying the e-beam current ($I_B$) while $V_B$ was kept constant at 1.5 kV. The integrated intensities are presented as a function of $I_B$ in Fig. 3(c). Varying $I_B$ in this range did not introduce any noticeable changes in peak shape or position of the RL band. This does not rule out a possible DAP recombination mechanism being responsible for the RL, however, as most donors and acceptors are expected to already exist in small-distance pairs due to the high H doping concentration. Fitting the data to a power law ($I_{CL} \propto I_B^k$) reveals a linear dependence of the UV band with the excitation density, having an exponent $k(UV) \approx 1$ within the experimental error of the measurement, for both the undoped and H-doped β-Ga$_2$O$_3$. This value is consistent with the fast decay dynamics of the STE emission, with a decay time of ~ 65 ns, in β-Ga$_2$O$_3$.[31] On the other hand, the RL exhibits a sub-linear dependence with $k(RL) = 0.6 \pm 0.2$. The sublinear dependence of the RL indicates CL saturation with increasing excitation density, which is generally observed due to saturation of a deep-level defect involved in the radiative recombination.[32] While we cannot unequivocally identify the specific acceptor defect responsible for the RL DAP transition, the logical candidate for the deep acceptor is a $V_{Ga}$-H complex, which has a lower formation energy than isolated $V_{Ga}$.[3] This $V_{Ga}$-H complex is highly stable with the activation energy for H dissociation predicted to be 3.4 eV.[3] The DAP emission energy is given by,[33,34]

$$h\nu(\text{DAP}) = E_g - (E_A + E_D) + \frac{e^2}{4\pi\epsilon r} \quad [1]$$

where $E_A$ and $E_D$ are the donor and acceptor binding energies, respectively. The last term accounts for the Coulombic interaction between the ionized donor and ionized acceptor with $r$ being their separation. The distance $r$ can be estimated as $r = \sqrt[3]{\frac{3}{4\pi N_D}}$ with $N_D$ being the donor concentration.[34] Using $N_D \approx 10^{17}$ cm$^{-3}$ for β-Ga$_2$O$_3$ doped with hydrogen under similar plasma conditions [10] and $E_D = 28$ meV for the measured donor binding energy, equation [1] yields $E_A \approx 2.88$ eV. This value is in excellent agreement with the predicted $V_{Ga}$-H$^I$ and $V_{Ga}$-H$^{II}$ complexes (for two inequivalent Ga sites) with energy levels of 3.02 and 2.82 eV above the VBM, respectively.[3] Accordingly, we attribute the RL to DAP recombination, in which the deep acceptor is a $V_{Ga}$-H complex and the donor is interstitial H. The passivation of compensating $V_{Ga}$ by H atoms is expected to contribute to the observed increase in the film conductivity as shown in Fig. 2(b).

## IV. CONCLUSIONS

The optical properties of hydrogen plasma doped β-$Ga_2O_3$ films were investigated and interpreted with regards to the corresponding chemical, structural and electrical characteristics. A RL band at 1.9 eV was identified and ascribed to the highly H-doped near-surface region of the film. The emergence of the RL was accompanied by an increase in the film electrical conductivity by an order of magnitude. Both temperature-dependent electrical conductivity and CL results indicated the presence of an H-related shallow donor with an ionization energy of 28 ± 4 meV. In view of the theoretically predicted behaviour of hydrogen in β-$Ga_2O_3$, the RL emission is attributed to radiative recombination involving the H-related shallow donor and $V_{Ga}$-H vacancy complexes.

**Acknowledgments**

This work was supported under Australian Research Council (ARC) Discovery Project funding scheme (project number DP150103317). This research was partly undertaken on the Soft X-ray Spectroscopy beamline at the Australian Synchrotron, Victoria, Australia. We would like to thank Dr Mark Lockrey for assistance with the CL measurements. We also thank Dr. Francois Jomard for conducting SIMS measurements.

**References**

[1]   S. J. Pearton, J. Yang, P. H. Cary IV, F. Ren, J. Kim, M. J. Tadjer, and M. A. Mastro, Appl. Phys. Rev. **5**, 011301 (2018).

[2]   M. Higashiwaki, K. Sasaki, H. Murakami, Y. Kumagai, A. Koukitu, A. Kuramata, T. Masui, and S. Yamakoshi, Semicond. Sci. Technol. **31**, 034001 (2016).

[3]     J. B. Varley, H. Peelaers, A. Janotti, and C. G. Van de Walle, J. Phys.: Condens. Matter **23**, 334212 (2011).

[4]     J. B. Varley, J. R. Weber, A. Janotti, and C. G. Van de Walle, Appl. Phys. Lett. **97**, 142106 (2010).

[5]     H. Gao, S. Muralidharan, N. Pronin, M. R. Karim, S. M. White, T. Asel, G. Foster, S. Krishnamoorthy, S. Rajan, L. R. Cao *et al.*, Appl. Phys. Lett. **112**, 242102 (2018).

[6]     P. Weiser, M. Stavola, W. B. Fowler, Y. Qin, and S. Pearton, Appl. Phys. Lett. **112**, 232104 (2018).

[7]     Y. Qin, M. Stavola, W. B. Fowler, P. Weiser, and S. J. Pearton, ECS J. Solid State Sci. Technol. **8**, Q3103 (2019).

[8]     J. E. N. Swallow, J. B. Varley, L. A. H. Jones, J. T. Gibbon, L. F. J. Piper, V. R. Dhanak, and T. D. Veal, APL Materials **7**, 022528, 022528 (2019).

[9]     S. Nakagomi, T. Sai, and Y. Kokubun, Sensor Actual B-Chem **187**, 413 (2013).

[10]    A. Y. Polyakov, I. H. Lee, N. B. Smirnov, E. B. Yakimov, I. V. Shchemerov, A. V. Chernykh, A. I. Kochkova, A. A. Vasilev, F. Ren, P. H. Carey *et al.*, Appl. Phys. Lett. **115**, 032101 (2019).

[11]    T. T. Huynh, L. L. C. Lem, A. Kuramata, M. R. Phillips, and C. Ton-That, Phys. Rev. Mater. **2**, 105203 (2018).

[12]    L. Binet and D. Gourier, J. Phys. Chem. Solids **59**, 1241 (1998).

[13]    X. Zhou, F. Heigl, J. Ko, M. Murphy, J. Zhou, T. Regier, R. Blyth, and T. Sham, Phys. Rev. B **75**, 125303 (2007).

[14]    T. Gake, Y. Kumagai, and F. Oba, Phys. Rev. Mater. **3**, 044603 (2019).

[15]    J. Varley, A. Janotti, C. Franchini, and C. G. Van de Walle, Phys.Rev.B. **85**, 081109 (2012).

[16] B. E. Kananen, N. C. Giles, L. E. Halliburton, G. K. Foundos, K. B. Chang, and K. T. Stevens, Journal of Applied Physics **122**, 6, 215703 (2017).

[17] J. B. Varley, A. Janotti, C. Franchini, and C. G. Van de Walle, Phys. Rev. B. **85**, 081109 (2012).

[18] K. Shimamura, E. G. Víllora, T. Ujiie, and K. Aoki, Appl. Phys. Lett. **92**, 201914 (2008).

[19] G. Pozina, M. Forsberg, M. Kaliteevski, and C. Hemmingsson, Sci. Rep. **7**, 42132 (2017).

[20] F. H. Teherani, D. J. Rogers, V. E. Sandana, P. Bove, C. Ton-That, L. L. C. Lem, E. Chikoidze, M. Neumann-Spallart, Y. Dumont, T. Huynh *et al.*, Proc. SPIE 10105, R1 (2017).

[21] J. J. Shi, H. W. Liang, X. C. Xia, Z. Li, Z. Long, H. Q. Zhang, and Y. Liu, J. Mater. Sci. **54**, 11111 (2019).

[22] F. P. Yu, S. L. Ou, and D. S. Wuu, Opt. Mater. Express **5**, 1240 (2015).

[23] S. Rafique, L. Han, and H. P. Zhao, Phys. Status Solidi A **213**, 1002 (2016).

[24] C. Kranert, C. Sturm, R. Schmidt-Grund, and M. Grundmann, Sci. Rep. **6**, 35964, 35964 (2016).

[25] M. Kneiss, A. Hassa, D. Splith, C. Sturm, H. von Wenckstern, T. Schultz, N. Koch, M. Lorenz, and M. Grundmann, APL Materials **7**, 022516, 022516 (2019).

[26] S. Ahn, F. Ren, E. Patrick, M. E. Law, S. J. Pearton, and A. Kuramata, Appl. Phys. Lett. **109**, 242108 (2016).

[27] J. Z. Sheng, E. J. Park, B. Shong, and J. S. Park, ACS Appl. Mater. Interfaces **9**, 23934 (2017).

[28] A. T. Neal, S. Mou, S. Rafique, H. P. Zhao, E. Ahmadi, J. S. Speck, K. T. Stevens, J. D. Blevins, D. B. Thomson, N. Moser *et al.*, Appl. Phys. Lett. **113**, 062101, 062101 (2018).

[29]  N. Moser, J. McCandless, A. Crespo, K. Leedy, A. Green, A. Neal, S. Mou, E. Ahmadi, J. Speck, K. Chabak *et al.*, IEEE Electron Device Lett. **38**, 775 (2017).

[30]  P. King, I. McKenzie, and T. Veal, Appl. Phys. Lett. **96**, 062110 (2010).

[31]  S. Yamaoka, Y. Furukawa, and M. Nakayama, Phys. Rev. B **95**, 094304, 094304 (2017).

[32]  C. Ton-That, L. Weston, and M. R. Phillips, Phys. Rev. B **86**, 115205, 115205 (2012).

[33]  D. G. Thomas, J. J. Hopfield, and W. M. Augustyniak, Phys. Rev. **140**, A202 (1965).

[34]  K. Thonke, T. Gruber, N. Teofilov, R. Schonfelder, A. Waag, and R. Sauer, Physica B **308**, 945 (2001).